\documentstyle[prl,aps,psfig]{revtex}
\begin{document}
\twocolumn[\hsize\textwidth\columnwidth\hsize
           \csname @twocolumnfalse\endcsname
\title{Thermopower in superconductors}
\author{Jan Kol\'a\v cek$^1$, Pavel Lipavsk\'y$^{1,2}$}
\address{$^1$Institute of Physics, Academy of Sciences, 
Cukrovarnick\'a 10, 16253 Prague 6, Czech Republic}
\address{$^2$Faculty of Mathematics and Physics,
Charles University, Ke Karlovu 5, 12116 Prague 2, Czech Republic}
\maketitle
\begin{abstract}
The thermopower of superconductors measured via the magnetic 
flux in bimetallic loop is evaluated. It is shown that by 
a standard matching of the electrostatic potential, known as the
Bernoulli potential, one explains the experimentally observed 
amplitude and the divergence in the vicinity of the critical 
temperature. 
\end{abstract}
    \vskip2pc]

The thermopower is a widely used tool to study electronic 
properties of conductive materials. An exception are 
superconductors, where the supercurrent cancels any
diffusive current so that the zero net current or voltage are
observed. This feature is known from 1935 \cite{SG35} and 
since then there were a number of attempts to access 
diffusive currents in an indirect way.

Already in 1944 Ginzburg \cite{G44} noticed that in 
inhomogeneous systems like bimetallic loops, the 
counter-flowing supercurrent creates a magnetic flux. The 
boom in this field came thirty years later. In 1974 Garland 
and Van~Harlingen proposed a simple phenomenological theory 
\cite{GH74} and Gal'perin, Gurevich and Kozub published 
a microscopic treatment \cite{GGK74} based on the 
Boltzmann-type approach. These theories predicted fluxes
of similar amplitudes and temperature dependences.

In the same year Zavaritskii presented first experimental 
data \cite{Z74} and he was soon followed by others 
\cite{PGP75,F76,VHG78}. Experimental results were 
a surprise. Zavaritskii \cite{Z74} and Falco \cite{F76} 
observed the expected temperature dependence, but
Pegrum, Gu\'enault and Pickett 
\cite{PGP75} and Van~Harlingen and Garland \cite{VHG78} 
monitored a thermally induced magnetic flux by five orders 
of magnitude larger. Moreover, the theory predicts 
that close to $T_c$ the flux $\Phi$ diverges as
${d\Phi\over dT}\propto (T_{\rm c}-T)^{-1}$, while a steeper
divergence ${d\Phi\over dT}\propto (T_{\rm c}-T)^{-3/2}$ 
was observed \cite{PGP75,VHG78}. The experimental situation 
in the late 1970 is reviewed in Ref.~\cite{MBB77}.

The giant flux stimulated a number of theoretical studies
\cite{PG76,K76,AV76,K78,HG78,S79} that explored various 
additional components ranging from a trapped flux, over 
impurities, over interfaces, to an influence of supercurrent flow. 
Most of these ingredients bring only a minor correction to the
original prediction. It was speculated, that the only sizable 
contribution can come from the trapped flux, 
which increasingly leaks into the ring as the 
temperature approaches its critical value. 
All these speculations were terminated by measurements of 
Van~Harlingen, Heidel and Garland \cite{HHG80}. To avoid 
the penetration of the external magnetic field 
they used the toroidal geometry and convincingly demonstrated 
that the large magnetic flux with the ${d\Phi\over dT}\propto 
(T_{\rm c}-T)^{-3/2}$ 
divergence is a genuine effect. By comparing a number of samples 
they could conclude  that the flux is proportional to the thermopower in 
the normal state and therefore that it is indeed caused by the thermal 
diffusion of electrons.

The lack of at least a qualitative theory has discouraged 
further measurements in this direction and the thermopower
joined the family of puzzling transport properties in
superconductors. Alternative measurements of the thermopower
via the superconducting fountain effect \cite{F77} or the
charge imbalance in the conversion region \cite{MCH84} yield
theoretically expected values but they have a rather low 
accuracy. In result, the values of the thermoelectric 
coefficients of superconductors are still not available, what
contrasts with extensive data gathered for superconducting
materials above $T_{\rm c}$. 

New theoretical interest in the giant magnetic flux emerged 
after seventeen years. Marinescu and Overhauser \cite{MO97} 
have analyzed the theory of Gal'perin, Gurevich and Kozub 
\cite{GGK74} and concluded, that its failure indicates a 
conceptual mistake in the underlying Boltzmann type transport 
theory developed by Bardeen, Rickayzen and Tewordt 
\cite{BRT59}. They made an {\em ad hoc} modification of the transport 
theory by including the momentum exchange between the 
condensate and quasiparticles. With this modification, a good 
agreement between theory and experimental data was reached. 

The modified transport theory, however, is in conflict with 
other properties of superconductors as discussed
recently by Gal'perin, Gurevich, Kozub and 
Shelankov \cite{GGKS02}.  From the time-reversal symmetry 
they showed that this theory predicts dissipative currents also 
in equilibrium systems with inhomogeneous chemical composition, 
i.e., in any real superconductor. 

Here we show that explanation of the thermopower does not
require any changes in the theory of transport 
in superconductors. The legitimate request of Marinescu and 
Overhauser to cover properly the balance of forces between 
the superconducting and normal electrons is naturally covered 
by the theory of electrostatic potential in superconductors known as the
Bernoulli potential. Unlike forces {\em ad hoc} added to the 
kinetic equation, forces derived from a scalar potential cannot result 
in an artificial dissipation.

Below we demonstrate that the giant magnetic flux can be described 
in a simple manner with the help of the Bernoulli potential. Our
approach parallels the textbook theory of thermopower in 
normal metals in that we evaluate the net current in the sample 
from the requirement of the electrostatic potential matching. 

\begin{figure}[h]  
  \centerline{\parbox[t]{8cm}{
  \psfig{figure=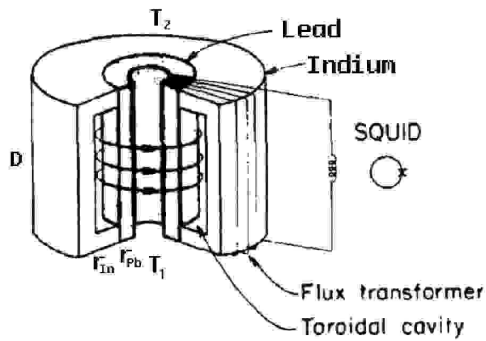,width=8cm}}}
\vskip 2pt
\caption{The toroidal sample of Van Harlingen {\em at al}
\protect\cite{HHG80}. The measured flux $\Phi$ is in the toroidal 
cavity.}
\label{f1}
\end{figure}

We will use notation related to the experimental setup of 
Van~Harlingen, Heidel and Garland \cite{HHG80} shown in 
Fig.~\ref{f1}.  The sample is a toroid with the internal cylinder 
from the Lead and the external one from the Indium. The magnetic 
flux $\Phi$ in question is restricted to the volume between 
cylinders, i.e., it is inside the toroidal cavity. The 
diamagnetic current $J$ corresponding to the flux $\Phi$ flows 
on the outer surface of Lead and the inner surface of Indium. 
The flux is linear in the current, 
\begin{equation}
\Phi=J{1\over 2\pi}\mu_0D
\ln{r_{\rm In}\over r_{\rm Pb}}, 
\label{eqPhi}
\end{equation}
where $D$ is the length of the sample, 
and $r_{\rm In,Pb}$ are the radii which enclose the flux.

The experiment is aimed to measure the transport coefficient
$L_T$ which determines the diffusive electric current 
${\bf j}_{\rm dif}=-L_T\nabla T$ caused by the temperature
gradient $\nabla T$. As already mentioned, ${\bf j}_{\rm dif}$ is 
not observable since it is cancelled by the supercurrent ${\bf j}
=-{\bf j}_{\rm dif}$. The theory of Garland and Van~Harlingen
\cite{GH74} uses the London gauge to find the vector potential 
${\bf A}=-\mu_0\lambda^2{\bf j}$, where $\lambda$ is the London 
penetration depth. The flux is then an integral along the 
bimetallic loop, $\Phi_{\rm G}=\oint d{\bf rA}=-\mu_0\oint 
d{\bf r}\lambda^2L_T\nabla T$. 

The London gauge is justified only for small fluxes. The data
from \cite{HHG80} show, however, that the flux is large so that 
it is of form $\Phi=N\Phi_0+\Phi_{\rm G}$, where $N$ is an 
integer quantum number of the superconducting condensate 
and $\Phi_0$ is the elementary flux. Estimates \cite{HHG80} 
indicate that $\Phi_{\rm G}\ll\Phi_0$, therefore to understand 
the giant flux, we have to find which state $N$ is the most 
favorable for the system with the imposed temperature 
gradient. Measured values of $N$ in \cite{HHG80} range up to 
250. These values are sufficiently large for the classical 
approximation, where $N$ is treated as a continuous variable. 
Accordingly, we will not assume quantum restrictions of the flux 
$\Phi$. 

The mechanism by which the flux arises is as follows. The 
diffusive current generates magnetic field, which is screened
by the counterflowing supercurrent. In the surface layer of the 
London penetration depth thickness, the cancellation is not 
complete. Accordingly, the supercurrent density is a sum of the 
counter-flow $-{\bf j}_{\rm dif}$ and a missing counter-flow 
$j{\rm e}^{-x/\lambda}$, where $x$ is a distance from the surface 
enclosing the cavity. We call it a diamagnetic current, as 
it screens the bulk of superconductor from the magnetic field, 
which is present in the toroidal cavity. 
 
Our aim is to find amplitudes $j$ in the Indium and the Lead. 
The total current 
$J=\int d^2{\bf r}({\bf j}+{\bf j}_{\rm dif})$ is the 
integral of the current density across each cylinder, i.e.,
$J=2\pi r_{\rm In}\lambda_{\rm In}j_{\rm In}$, 
and due to continuity condition also
$J=-2\pi r_{\rm Pb}\lambda_{\rm Pb}j_{\rm Pb}$. Since
$\lambda$ depends on the temperature, the surface values of 
the diamagnetic current densities $j_{\rm In,Pb}$ change along 
the temperature gradient, while the product $\lambda j$ stays
constant.  

Now we specify the condition for the total current $J$ from the
requirement of the scalar potential matching. As was observed by 
Bok and Klein \cite{BK68} and with a higher precision by Morris 
and Brown \cite{BM68,MB71}, current in the superconductor 
induces perpendicular electric field. It is well approximated
by the electrostatic potential of Bernoulli type \cite{VS64}
\begin{equation}
\varphi={n_s\over n}{1\over 2}mv^2,
\label{e1}
\end{equation}
where $n_s$ is the density of superconducting electrons. 
The velocity of the superconducting electrons at the surface 
is given by the current density,
\begin{equation}
v={{\bf j}\over en_s}={L_T\nabla T\over en_s}
\pm{J\over 2\pi r\lambda en_s},
\label{e2}
\end{equation}
where plus applies for the Indium and minus for the Lead, in which 
the total current flows in opposite direction. The first term is due 
to the compensating supercurrent $-{\bf j}_{\rm diff}$, the 
second term is caused by the diamagnetic current $j$.

The electrostatic potential has to be continuous, therefore the 
potential differences created by the temperature gradient in the 
Lead and the Indium has to be equal, 
\begin{equation}
\varphi_{\rm Pb}(T_2)-\varphi_{\rm Pb}(T_1)=
\varphi_{\rm In}(T_2)-\varphi_{\rm In}(T_1).
\label{e3}
\end{equation}
This is the central equation in our approach. From the set
(\ref{e1}-\ref{e3}) one can directly evaluate the current $J$ and 
the magnetic flux (\ref{eqPhi}). 

Condition (\ref{e3}) is a simple quadratic equation, which 
includes material parameters of both, the Lead and Indium. 
The experiment \cite{HHG80} explores temperatures close to critical 
temperature of Indium $T_c^{\rm In} =3.4$~K, therefore only a
small fraction of electrons remain superconducting  
$n_s^{\rm In}\ll n^{\rm In}$ in the Indium arm. Since
the critical temperature of Lead $T_c^{\rm Pb} = 7.19$~K is 
considerably higher, the majority of electrons are superconducting, 
$n_s^{\rm Pb}\sim n^{\rm Pb}$, and consequently the difference of Bernoulli 
potential in the Lead is much smaller than the potential difference in the 
Indium. Briefly, the Lead effectively short-circuits 
ends of the Indium, so that (\ref{e3}) reduces to 
$\varphi_{\rm In}(T_2)=\varphi_{\rm In}(T_1)$ and material parameters 
of Lead drop out.

The second simplification follows from the relation between 
the superconducting density and the London penetration depth,
$\lambda^2=m/(\mu_0e^2n_s)$. This allows us to express the 
condition (\ref{e3}) on the potential as $v_{\rm In}(T_1)/
\lambda_{\rm In}(T_1)=-v_{\rm In}(T_2)/\lambda_{\rm In}(T_2)$. 
The net current is now a solution of a linear relation and the 
resulting magnetic flux reads
\begin{equation}
\Phi=-{1\over 2}\mu_0L_T(T_2-T_1)\left(
\lambda(T_1)+\lambda(T_2)\right)
r_{\rm In}\ln{r_{\rm In}\over r_{\rm Pb}}.
\label{e4}
\end{equation}
In this expression, the thermoelectric coefficient $L_T$ 
and the London penetration depth $\lambda$ are of the 
Indium. To make the expression compact we write $D\nabla T=T_2-T_1$.

\begin{figure}[h]  
  \centerline{\parbox[t]{15cm}{
  \psfig{figure=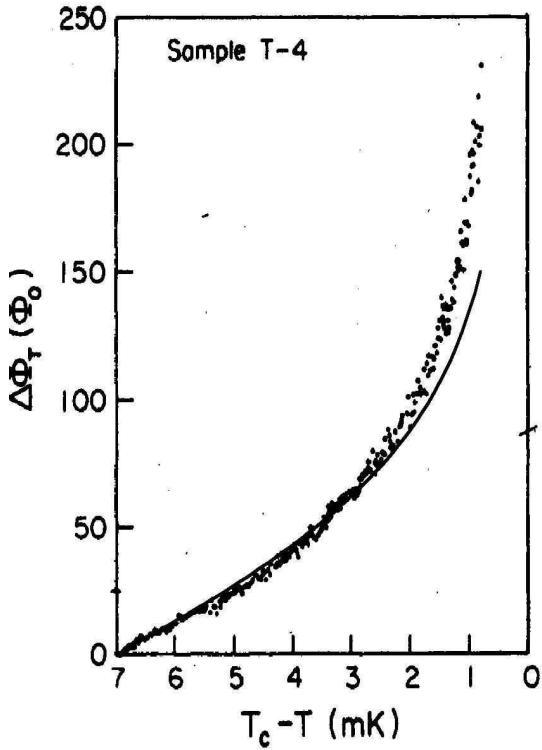,width=8cm}}}
\vskip 2pt
\caption{Thermally induced magnetic flux in a toroid. The dots 
are the experimental data of Van~Harlingen, Heidel and 
Garland \protect\cite{HHG80}, the line is according to 
formula (\ref{e4}) of this paper with $T_1=T_{\rm c}-7$mK and 
$T_2\equiv T$.}
\label{f2}
\end{figure}
In Fig.~\ref{f2} we compare experimental data of Van~Harlingen, 
Heidel and Garland \cite{HHG80} with formula (\ref{e4}). The
diameters of toroid are $r_{\rm In}=3$~mm and $r_{\rm In}=1
$~mm. Material parameters are the thermoelectric coefficient 
of the normal metal, $L_T=9.82~10^3$~A/Km, and the London 
penetration depth at the zero temperature, $\lambda_0=
400$~\AA. All the parameters are from Ref.~\cite{HHG80} as 
values for the sample T-4. The only open question is
the temperature dependence of the London penetration depth 
close to the critical temperature $T_{\rm c}^{\rm In}=3.4$~K. 
We use the asymptotic BCS
relation $\lambda=\lambda_0/\sqrt{2-2T/T_{\rm c}}$.

In the linear region, $T_1-T_2\ll T_{\rm c}-T_2$, the theory
agrees with the experimental data within experimental errors. 
Perhaps we should return to the original aim of the 
measurement and conclude that experimental data from 
\cite{HHG80} confirm that the thermoelectric coefficient 
$L_T$ close below $T_{\rm c}$ has the same value as close 
above. 

In the non-linear region the theory deviates from data. This is 
no surprise since the presented theory is locally linear in the 
temperature gradient. Moreover, additional non-linear effects 
are caused by the so called thermodynamic correction to the 
electrostatic potential \cite{R69,LKMB02}. We aim to discuss 
these corrections in a next paper. 

We should mention that formula (\ref{e4}) has been derived 
under a tacit assumption that a width $w_{\rm In}$ of the 
Indium cylinder is sufficiently larger than the London 
penetration depth $\lambda$. For $\lambda(T_2)\to w_{\rm In}$ 
the magnetic flux given by formula (\ref{e4}) approaches 
$\Phi\to{1\over 2}\Phi_{\rm n}$, where $\Phi_{\rm n}=-\mu_0
L_T(T_2-T_1)w_{\rm In}r_{\rm In}\ln{r_{\rm In}\over r_{\rm Pb}}$. 
The flux $\Phi_{\rm n}$ develops when the Indium makes transition 
to the normal (non-superconductive) state, while the Lead remains
superconducting. The factor ${1\over 2}$ results from the
unrestricted integration of the diamagnetic current into the bulk
of Indium valid only for $\lambda\ll w$. Assuming the upper 
integration limit, $J=\int_0^w dx\,j$, one finds that 
$\Phi\to\Phi_{\rm n}$. Unfortunately,
details of the flux in the narrow vicinity of $T_{\rm c}$ have 
not been measured. One can merely speculate that $\Phi_{\rm n}$ 
is the upper limit of the diverging flux $\Phi$.

Finally, we want to clarify the simple potential matching used
above. First, the potential has to match across the whole sample
while equation (\ref{e3}) was obtained by matching only at the
inner surface. At the outer surface the Bernoulli potential 
(\ref{e1}) is zero everywhere so that the matching is clearly 
satisfied. The potential profile between the inner and outer
surfaces is nontrivial since the current profile is complicated
by itself. Indeed, the screening currents in the Indium and in
the Lead spread over different London penetration depths, and 
they have to match across the whole interface with the current
continuity satisfied at each point. We plan to evaluate the 
current and potential profiles in future. The present theory 
is based on our believe that the matching at the surface points 
of the interface is sufficient.

Second, we have ignored the role of the flat pieces in the
upper and lower end of sample. In these pieces, the temperature 
gradient is absent, nevertheless, the potential difference 
across each piece is nonzero since the current density at the 
matching corner to the Lead is higher than the current density 
at the Indium corner. Sending $\nabla T$ to zero in (\ref{e2}) 
and using obtained velocities in (\ref{e1}), we find that the 
potential differencies across the upper and the lower flat 
pieces are identical and thus cancel.

Third, the Bernoulli potential (\ref{e1}) is the simplest
approximation of the electrostatic potential. Why we ignore
more sophisticated potentials that include the thermodynamic
corrections \cite{R69} and non-local corrections due to the
finite Ginzburg-Landau coherence length \cite{LMKMBS04}? Both 
these corrections result in a surface dipole \cite{LKM01} 
which makes the potential matching more complex. On the other 
hand, with the surface dipole there also appears a dipole at 
the interface of Indium and Lead. We expect that these dipoles 
tend to cancel in the final matching condition.

In conclusion, we would like to encourage measurements of 
thermoelectric coefficients in superconductors. Since the 
detection of magnetic fluxes is extremely sensitive and
fluxes $10^{-3}\Phi_0$ can be conveniently monitored, it 
should be possible to access $L_T$ in a wider temperature 
region, not merely few mK's bellow $T_{\rm c}$. 

Far from $T_{\rm c}$ the present theory is not valid, since 
the magnetic flux becomes small and it has to exhibit the 
quantization. Fluxes smaller than the elementary flux
are covered by the former theory \cite{GH74,GGK74}, as 
confirmed by Zavaritskii \cite{Z74} and Falco \cite{F76}, who 
monitored fluxes of the order of $10^{-2}\Phi_0$ and 
$10^{-1}\Phi_0$, respectively. 

Interesting features might appear at the 
intermediate region, where fluxes are comparable to the 
elementary flux. For instance, the above discussed sample has 
the classical estimate of thermally induced flux equal to 
$1\Phi_0/$mK at the temperature $T=T_{\rm c}-65$~mK. It 
should be thus in an access of experiment to observe whether
the flux increases in steps $\Phi_0$ or smoothly.

\vskip 0.5cm
We are grateful to Van Harlingen for a kind permission
to reproduce their figures.
This work was supported by M\v{S}MT program Kontakt 
ME601 and GA\v{C}R 202/03/0410, GAAV A1010312 grants. 
The European ESF program VORTEX is also acknowledged.

\end{document}